\documentclass[%
 reprint,
superscriptaddress,
 amsmath,amssymb,
 aps,
]{revtex4-2}

\pdfoutput=1

\usepackage[utf8]{inputenc}
\usepackage[T1]{fontenc}

\usepackage{graphicx}
\usepackage{dcolumn}
\usepackage{bm}
\usepackage{color}
\usepackage{graphics}
\usepackage{subfigure}
\usepackage{pdftexcmds}
\usepackage{ifpdf}
\usepackage{braket}
\usepackage{dsfont}

\usepackage{url}

\usepackage{capt-of}

 \newcommand{\bc}{\begin{center}}
 \newcommand{\ec}{\end{center}}
 
\newcommand{\be}{\begin{equation}}
\newcommand{\ee}{\end{equation}}

\newcommand{\bea}{\begin{eqnarray}}
\newcommand{\eea}{\end{eqnarray}}

\newcommand{\bd}{\begin{displaymath}}
\newcommand{\ed}{\end{displaymath}}

\newcommand{\CrNbS}{$\mathrm{CrNb}_3\mathrm{S}_6$}
\newcommand{\hcrit}{h_{\mathrm{c}}}

\newcommand{\he}{\hat{e}}

\newcommand{\hx}{\hat{x}}
\newcommand{\hy}{\hat{y}}
\newcommand{\hz}{\hat{z}}

\newcommand{\hB}{\hat{B}}
\newcommand{\vtau}{\vec{\tau}}
\newcommand{\vc}{\vec{c}}
\newcommand{\ve}{\vec{e}}
\newcommand{\vh}{\vec{h}}
\newcommand{\vm}{\vec{m}}

\newcommand{\vB}{\vec{B}}
\newcommand{\vT}{\vec{T}}
\newcommand{\Be}{\vec{B}_{\mathrm{eff}}}
\newcommand{\bef}{\vec{b}_{\mathrm{eff}}}

\newcommand{\atan}{\,\mathrm{atan}\,}

\newcommand{\e}{\text{e}}

\newcommand{\pmin}{p_\text{min}}
\newcommand{\pmax}{p_\text{max}}

\begin{document}

\title{Helicity in monoaxial chiral magnets}

\author{Victor Laliena}
\affiliation{
Department of Applied Mathematics and Institute of Mathematics and Applications (IUMA), University of Zaragoza,
C/ Mar\'ia de Luna 3, 50018 Zaragoza, Spain
}

\author{Diego Giron\'es-Maga\~na}
\affiliation{
Aragon Nanoscience and Materials Institute (CSIC-University of Zaragoza) and Condensed Matter Physics Department, University of Zaragoza, C/ Pedro Cerbuna 12, 50009 Zaragoza, Spain
}

\author{Javier Campo}
\affiliation{
Aragon Nanoscience and Materials Institute (CSIC-University of Zaragoza) and Condensed Matter Physics Department, University of Zaragoza, C/ Pedro Cerbuna 12, 50009 Zaragoza, Spain
}
\affiliation{International Institute for Sustainability with Knotted and Chiral Meta Matter  (Visiting Professor), 2-313 Kagamiyama, Higashi-Hiroshima, Hiroshima, 739-0046, Japan}  

\date{\today}

\begin{abstract}
The equilibrium state of a monoaxial chiral magnet surrounded by a non-magnetic medium (such as air or vacuum) is a helical texture characterized by a single, well-defined wave vector. No metastable states have ever been observed in such systems. Recently, however, it was demonstrated that when a chiral magnet is in close contact with two uniaxial ferromagnets, a large number of metastable helical states emerge in addition to the equilibrium state [Phys. Rev. B 109, 214424]. These helical states are distinguished by their wave number (helicity). In the present work, we elucidate the topological origin of the stabilization of these states—a mechanism we term dynamical topological protection—and investigate their static and dynamic properties. We find that, as a consequence of this dynamical topological protection, the winding number of the helical states remains constant under the application of sufficiently weak magnetic fields and polarized electric currents. Furthermore, when a polarized current is applied to a metastable helical state, a static configuration is reached. This state retains the original winding number, but its winding number density becomes non-homogeneously distributed, concentrating near the interface with one of the ferromagnets. The dynamic response to sufficiently large magnetic fields and currents provides mechanisms to switch between different helical states. Since the magnetic properties depend on helicity, these metastable helical states are highly promising for applications in spintronics and magnonics.
\end{abstract}

\maketitle

\section{Introduction}

In this work, we study the metastable helical and conical states of monoaxial chiral magnets, as well as the dynamic processes that enable switching among them via applied magnetic fields and polarized electric currents. The noncollinearity of the magnetization characteristic of these states makes them highly interesting, both from a fundamental physics perspective and in terms of their potential applications in spintronics and magnonics. \cite{Back2020,Barman2021,Mruczkiewicz2021,Vedmedenko2020}.

The main feature of monoaxial chiral magnets is the presence of a relatively weak Dzyaloshinskii-Moriya interaction (DMI) that acts only along one specific direction, which coincides with a particular crystallographic axis. We call this direction the \textit{chiral axis}. In addition, these materials show a strong uniaxial anisotropy of easy-plane type, with the anisotropy axis parallel to the chiral axis. The competition between the DMI and the ferromagnetic symmetric exchange interaction is the origin of the noncollinear textures. At low temperature and in the absence of an applied field, the equilibrium state exhibits a helical magnetic texture where the magnetization lies within the easy plane, the propagation vector aligns with the chiral axis, and the wave number is determined by the relative intensities of the DMI and the symmetric exchange interaction. The DMI induces magnetic chirality because the direction of the propagation vector is fixed by energy minimization, implying that helical textures with the opposite propagation vector have higher energy. Therefore, the equilibrium state is not invariant under a reflection about the easy plane
 
If a magnetic field is applied along the chiral axis, the equilibrium state becomes a conical texture with the same propagation vector as the zero-field helical state. As the field strength increases, the magnetization component along the chiral axis grows until a uniform state, with the magnetization fully aligned with the field, is eventually reached. If the field is applied perpendicular to the chiral axis, the helical texture deforms and acquires solitonic features, resulting in a non-homogeneous energy distribution throughout the material—a texture known as a chiral soliton lattice. When the field strength surpasses a critical value, the equilibrium state once again becomes uniform, with the magnetization aligned with the field. Finally, if the field is applied along a generic direction—neither parallel nor perpendicular to the chiral axis—the magnetic texture exhibits a hybrid character that combines solitonic and conical features.
As expected, this noncollinear texture is replaced by a uniform state aligned with the field if the field intensity is sufficiently large
 \cite{Miyadai1983,Togawa2012,Laliena2016a,Laliena2016b,Laliena2017a,Ghimire2013,Chapman2014,
Tsuruta2016,Yonemura2017,Clements2017,Clements2018,Osorio2023}. 
Even at high magnetic fields, noncollinear textures persist, since the uniform state can host metastable isolated solitons \cite{Laliena2020}. The archetypical monoaxial helimagnet is CrNb$_3$S$_6$ \cite{Moriya1982,Kishine2005}, but there are many others, for instance CrTa$_3$S$_6$,  MnNb$_3$S$_6$, CuB$_2$O$_4$, CuCsCl$_3$, Yb(Ni$_{1-x}$Cu$_x$)$_3$Al$_9$, or Ba$_2$CuGe$_2$O$_7$
 \cite{Kousaka2016,Karna2019,Roessli2001,Adachi1980,Ohara2014,Matsumura2017,Zheludev1997}.

The magnetic states that have received the most attention for a long time are topologically non-trivial textures, such as the skyrmions found in cubic chiral magnets. Nevertheless, there has recently been a revival of interest in helical and conical states. In cubic chiral magnets, these states occupy a larger region of the phase diagram and are thus more easily stabilized, offering a clear advantage. In contrast to monoaxial chiral magnets, the equilibrium helical states in cubic chiral magnets are degenerate, since the helix wave vector can point along different equivalent crystallographic directions selected by the cubic anisotropy. It has been shown that the wave vector direction can be controlled via electrical means. Consequently, helical states with different wave vectors can be used to store and manipulate information \cite{Masell2020prb}. The direction of the helix wave vector can also be changed through thermal currents \cite{Yasin2023}. Therefore, the orientation of helical stripes may serve as a building block for devices in classical and unconventional computing \cite{Bechler2023}

In monoaxial chiral magnets, the degeneracy of the helical state is absent since the helix wave vector points along the chiral axis in the direction determined by the DMI. However, in a magnet without boundaries—either because the system occupies the whole space or because periodic boundary conditions along the chiral axis are imposed—there are many metastable helical states differing by the helix wave number \cite{Laliena2023,Laliena2018a} (similar states also exist in cubic chiral magnets without boundaries \cite{Laliena2017b}). It is possible to switch between these helical states by applying magnetic fields and electric currents along the chiral axis \cite{Laliena2023}. It has been shown in Ref.~\onlinecite{Laliena2024} that in a real magnet, which has boundaries, a large number of metastable helical states also exists if a parallelepiped-shaped monoaxial chiral magnet has two slabs of a ferromagnetic material attached to its two faces perpendicular to the chiral axis, as in Fig.~\ref{fig:composite_magnet}. For brevity, we call this system a composite magnet. The metastable helical states of composite magnets may also serve as building blocks for computing devices, provided that there are means to switch between them.

It is convenient to introduce some terminology here. The wave number of a helical or conical state, in convenient units, is called the \textit{helicity}, and coincides with the winding number density of the state (see Sec.~\ref{sec:topo}). For more general states, the \textit{winding number density} plays the role of helicity.

In this work, we unveil the topological origin of the stabilization of metastable helical states, which relies on the combined properties of the chiral magnet and the ferromagnet, as well as the geometry shown in Fig. \ref{fig:composite_magnet}. Consequently, the stabilization mechanism is absent when the chiral magnet is in contact solely with non-magnetic media, which explains why no metastable helical state has ever been observed in chiral magnets. We also demonstrate that switching between these metastable helical states can be achieved by applying magnetic fields and spin-polarized electric currents. To this end, we investigate their static and dynamic properties in detail. Since the magnetization cannot be obtained analytically under an applied field or current, we resort to numerical simulations, which are carried out using the computational codes described below.

We find that the effect of an applied field on the helical states of the composite magnet is similar to that in the chiral magnet without boundaries \cite{Laliena2023}. The behavior under an applied current is more intriguing. In the chiral magnet without boundaries, a helical state subjected to a polarized electric current reaches a steady motion state in which the magnetization moves rigidly with a velocity proportional to the current density \cite{Laliena2023}. In a composite magnet, this steady state is clearly impossible. We show that, in this case, the magnetization reaches a static state whose winding number density is equal to the helicity of the initial helical state. A similar behavior is obtained if the current and the field are simultaneously applied.

The rest of the paper is organized as follows. Section~\ref{sec:model} describes the magnetic system and establishes the notation and physical parameters. Section~\ref{sec:helical} examines the metastable helical states, while Section~\ref{sec:topo} provides the topological argument explaining their stability. The effect of an applied magnetic field on these helical states is addressed in Section~\ref{sec:conical}. Next, Section~\ref{sec:static} analyzes the static states reached under an electric current, with or without an external field, and Section~\ref{sec:switch} investigates dynamic processes relevant to controlling these helical states. Finally, Section~\ref{sec:conc} summarizes our conclusions.

\section{Composite chiral magnet \label{sec:model}}

\begin{figure}[t!]
\includegraphics[width=0.3\textwidth]{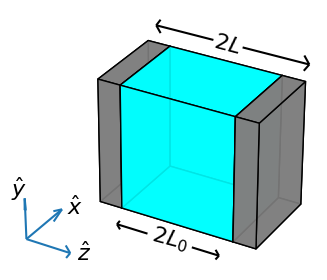}
\caption{Composite chiral magnet. The cyan region is occupied by a monoaxial chiral magnet, and the gray regions by two identical slabs of a uniaxial ferromagnet. The chiral axis is oriented along the $\hz$ direction. Although not reflected in the figure, the dimensions of the system along the $\hx$ and $\hy$ directions are assumed to be infinite.
\label{fig:composite_magnet}}
\end{figure}

Let $\{\hx,\hy,\hz\}$ be an orthonormal triad that defines a Cartesian coordinate system in space, with coordinates denoted by $(x,y,z)$. We consider a magnetic system of a rectangular parallelepiped shape which occupies a region of size $2L$ along the $\hz$ direction, so that $-L\leq z\leq L$. The dimensions of the system in the $\hx$ and $\hy$ directions are very large and thus are considered infinite. The system is inhomogeneous along the $\hz$ direction and consists of three homogeneous pieces: a monoaxial chiral magnet occupies the central region of size $2L_0$, that is, the region $-L_0\leq z\leq L_0$; the peripheral regions, $-L\leq z < -L_0$ and $L_0 < z \leq L$, are occupied by two similar uniaxial ferromagnets, as shown in Figure \ref{fig:composite_magnet}. The materials are oriented so that the chiral axis of the monoaxial chiral magnet is aligned with the $\hz$ axis, and the easy axis of each ferromagnet is aligned with the $\hx$ axis. The direction of the magnetization is given by the unit vector field $\vm$.

The chiral magnet is characterized by the following physical properties: the saturation magnetization, $M_c$; the stiffness constant, $A_c$, which measures the strength of the effective symmetric exchange interaction; the uniaxial DMI, whose strength is denoted by $D$; and the uniaxial anisotropy, which is of the easy-plane type, whose axis coincides with the chiral axis $\hz$, and whose energy per unit volume is given by the anisotropy constant $K_c<0$. The uniaxial ferromagnet has a saturation magnetization $M_u=\mu M_c$ and a stiffness constant $A_u=\rho A_c$, where $\mu$ and $\rho$ are positive dimensionless constants. The ferromagnet's easy axis runs along $\hx$, and its anisotropy constant is $K_u>0$.

It is convenient to introduce the characteristic functions $\chi_c(z)$ and $\chi_u(z)$, defined as $\chi_c(z)=1$ if $\vert{}z\vert{}\leq L_0$ and $\chi_c(z)=0$ otherwise, and as $\chi_u(z)=1$ if $L_0<\vert{}z\vert{}\leq L$ and $\chi_u(z)=0$ otherwise. The total energy of the system, $E$, in the presence of an external magnetic field $\vB$, is given by the volume integral of the energy density, $w$, which contains a nonlocal term, $w_m$, induced by the magnetostatic energy:
\be
\begin{gathered}
w=w_m + a_s A_c \sum_i\partial_i\vm\cdot\partial_i\vm + \chi_cD_c\vm\cdot(\hz\times\partial_z\vm) \\
-\chi_cK_c(\hz\cdot\vm)^2 - \chi_uK_u(\hx\cdot\vm)^2 - a_m M_c\vB\cdot\vm,
\end{gathered}
\ee
where $a_s$ and $a_m$ are the piecewise constant functions
\be
a_s = \chi_c + \rho\chi_u, \qquad a_m = \chi_c + \mu\chi_u.
\ee
In this work, only one-dimensional modulations along the $\hz$ axis are considered, and thus the magnetostatic energy density exhibits a local character, having the form of an easy-plane uniaxial anisotropy given by \cite{Hubert2008}
\be
w_m = \frac{\mu_0M_c^2}{2}\big(\hz\cdot\vm\big)^2a_m^2.
\ee
The contribution of the chiral magnet to $w_m$ is absorbed into the anisotropy constant $K_c$, so that $w_m=(\mu_0\mu^2M_c^2/2)(\hz\cdot\vm)^2\chi_u$.

We introduce the quantities $q_c$, $q_u$, and $q_m$, which have dimensions of inverse length,
\be
q_c=\frac{D_c}{2A_c}, \quad q_u^2 = \frac{K_u}{\rho A_c}, \quad q_m^2 =\frac{\mu^2}{\rho}\, \frac{\mu_0M_c^2}{2 A_c}, 
\ee
 and the dimensionless parameters 
 \be
 \kappa_c=\frac{4A_cK_c}{D_c^2}, \quad \kappa_u = \frac{q_u^2}{q_c^2}, \quad \kappa_m = \frac{q_m^2}{q_c^2}.
 \ee
 
To discuss the conditions that $\vm$ has to satisfy at the internal interfaces and the external boundaries, we introduce the vector field
\be
\vc = a_s \partial_z \vm - \chi_c q_c\hz\times\vm. \label{eq:vc}
\ee
Both $\vm$ and $\vc$ have to be continuous at the internal interfaces, $z=\pm L_0$, and $\vc$ has to vanish at the external boundaries, $z=\pm L$ \cite{Laliena2024}.  In the present case, the conditions at the external boundaries are simply Neumann boundary conditions.

The effective field is given by the functional derivative $\Be=-(1/a_sM_c)\delta E/\delta\vm$. If $\vm$ satisfies the conditions described above (namely, the continuity of $\vm$ and $\vc$), integration by parts can be applied to obtain the functional derivative in the standard way. For one-dimensional modulations along $\hz$, we obtain $\Be=(2A_cq_c^2/M_c)\bef$, with
\be
\begin{gathered}
\bef =  \frac{1}{a_mq_c^2}\,\vc^{ \,\,\prime} - \frac{1}{q_c}\chi_c\hz\times\vm^{\,\prime} 
+ \frac{\rho\kappa_u}{\mu} \chi_u(\hx\cdot\vm)\hx   \\[2pt]
+ \Big(\kappa_c \chi_c+ \frac{\rho \kappa_m}{\mu}\chi_u\Big)(\hz\cdot\vm)\hz 
+ h\hB,
\label{eq:beff}
\end{gathered}
\ee
where $h=2A_cM_cB/D^2$ and $\hB=\vB/B$, and the prime denotes the derivative with respect to the coordinate $z$. We assume that $\vm$ is a continuous, piecewise smooth function of $z$. The first term on the right-hand side of equation (\ref{eq:beff}) is proportional to the derivative of $\vc$. Were this vector field not continuous at the interfaces, its derivative would induce terms proportional to Dirac delta functions in the effective field, which are not present in the energy density. The continuity of $\vc$ ensures the absence of Dirac delta terms in the effective field.
 
The magnetization dynamics is governed by the Landau-Lifshitz-Gilbert (LLG) equation,
 \be
 \partial_t \vm = \gamma\Be\times\vm + \alpha\vm\times\partial_t\vm + \vT,
 \ee
 where $\alpha$ is the Gilbert damping parameter, $\gamma$ is the gyromagnetic ratio, and $\vT$ is a torque delivered by some external agent not included in the system energy. We consider torques delivered by spin-polarized electric currents applied along $\hz$, which take the Zhang-Li form \cite{Zhang2004}:
 \be
 \vT = -\tau_0\big(\vm\times\vm^{\,\prime} + \beta \vm^{\,\prime}\big)\times\vm,
 \ee
with $\tau_0=\mu_\mathrm{B}Pj/M_ca_m|e|(1+\beta^2)$, where $\mu_\mathrm{B}$ is the Bohr magneton, $e$ is the electron charge, $j\hz$ is the electric current density, $P$ is the degree of polarization of the current, and $\beta$ is the non-adiabaticity parameter. The derivatives in the Zhang-Li torque act only on the continuous, piecewise smooth vector field $\vm$, and thus the torque does not contribute to either the interface or the boundary conditions.

The Gilbert damping parameter $\alpha$ and the non-adiabaticity parameter $\beta$ are different in different materials, and thus in this work they should be modeled as piecewise constant functions, taking different values in the chiral magnet and in the ferromagnet. However, to avoid parameter proliferation, we adopt the widely used values $\alpha=0.01$ and $\beta=0.02$ in both materials.
 
It is convenient to cast the LLG equation into the form
\be
\partial_t \vm = \omega_c \Big(\bef+\vtau + \alpha \vm\times(\bef+\vtau)\Big)\times\vm,
\ee
where $\omega_c=2A_cq_c^2\gamma/M_c(1+\alpha^2)$ has the dimensions of frequency and
\be
 \vtau = -\frac{v}{a_mq_c}\big(\vm\times\vm^{\,\prime} + \beta \vm^{\,\prime}\big),
 \ee
where $v=M_c\tau_0/2A_cq_c\gamma$ is dimensionless.

Static states, with $\partial_t \vm=0$, are solutions of the static equation
\be
\big(\bef + \vtau\big)\times\vm = 0. \label{eq:static}
\ee
The explicit form of the static equation, parameterizing $\vm$ in polar coordinates, is shown in appendix \ref{app:static}, in which the stability conditions of the static states are also studied.

At zero field and zero current, this system possesses metastable helical states that can be obtained exactly \cite{Laliena2024}. To investigate the behavior of these states under the action of applied currents and fields, we have to resort to numerical methods. We developed two finite-difference algorithms: one to solve the LLG equation, and another to solve the static equation and analyze its stability conditions. These algorithms were implemented in two separate computational codes. For brevity, we refer to the code solving the LLG equation as the \textit{dynamic code}, and the code solving the static equation as the \textit{static code}.

For the analysis carried out in this work, we adopt the material parameters listed in Table \ref{tab:params}. The parameters for the chiral magnet are those of \CrNbS, the most widely studied monoaxial chiral magnet. The parameters for the uniaxial ferromagnet correspond roughly to those of hard magnets used in memory devices, such as FePt. As previously stated, we set $\alpha=0.01$ and $\beta=0.02$ for both materials. The external parameters (magnetic field and electric current strengths) are given by the dimensionless variables $h$ and $v$, respectively. For the physical parameters in Table \ref{tab:params}, a unit of $h$ corresponds to $0.37$~T, and, assuming a fully spin-polarized current ($P=1$), a unit of $v$ corresponds to $1.12 \times 10^{12}$~A/m$^2$.We define $1/q_c \approx 7.7$~nm as the unit of length and $1/\omega_c \approx 0.015$~ns as the unit of time. The system sizes considered are $q_c L_0 = 100$ and $q_c L = 140$, which correspond to $L_0 \approx 770$~nm and $L \approx 1078$~nm, respectively.

\begin{table}[t!]
\begingroup
\setlength{\tabcolsep}{4pt} 
\renewcommand{\arraystretch}{1.5} 
\begin{center}
\begin{tabular}{cccc}
\hline\hline
$M_c$ (A/m) & $A_c$ (J/m) & $D_c$ (J/m$^2$) & $K_c$ (J/m$^3$)  \\ [1pt]
\hline
1.29$\cdot$10$^5$ & 
1.42$\cdot$10$^{-12}$ & 
3.69$\cdot$10$^{-4}$ & 
-1.24$\cdot$10$^{5}$
\\
\hline\hline
\end{tabular}
\end{center}
\begin{center}
\begin{tabular}{ccc}
\hline\hline
$M_u$ (A/m) & $A_u$ (J/m) &  $K_u$ (J/m$^3$)  \\ [1pt]
\hline
1.29$\cdot$10$^6$ & 
9.94$\cdot$10$^{-12}$ &  
4.2$\cdot$10$^{6}$
\\
\hline\hline
\end{tabular}
\end{center}
\begin{center}
\begin{tabular}{ccccccc}
\hline\hline
$\mu$ & $\rho$ & $\kappa_c$ & $\kappa_u$ & $\kappa_m$ & $q_c$ (nm$^{-1}$) & $\omega_c$ (GHz)\\ [1pt]
\hline
10.0 & 7.0 & -5.17 & 25.0 & 6.23 & 0.13 & 65.4  \\
\hline\hline
\end{tabular}
\end{center}
\endgroup
\captionof{table}{Material parameters used in this work.
\label{tab:params}}
\end{table}

\section{Helical states \label{sec:helical}}

In the absence of an applied field and current, the system possesses multiple metastable helical states differing by their helicity \cite{Laliena2024}, which are of the form
$
\vm = \cos\varphi\,\hx + \sin\varphi\,\hy. 
$
The angle $\varphi$ is a function of $z$ given by
\be
\varphi(z) = \left\{
\begin{array}{cc}
-\sigma_p\,\varphi_0(-z), & \;-L<z<-L_0, \\[4pt]
pq_cz + n\pi, & \;-L_0<z<L_0, \\[4pt]
\sigma_p\,\varphi_0(z), & \;L_0<z<L, 
\end{array}
\right.
\label{eq:phi}
\ee
where $\sigma_p=1$ if $p\geq1$ and $\sigma_p=-1$ if $p<1$, and $n$ can only take the values 0 or 1. 
The function $\varphi_0(z)$ is an appropriate solution to the sine-Gordon equation, and describes a piece of a domain wall on each ferromagnet slab (see Ref. \onlinecite{Laliena2024} for details). These states have the symmetry
\be
\hx\cdot\vm(-z)=\hx\cdot\vm(z), \quad \hy\cdot\vm(-z)=-\hy\cdot\vm(z),
\ee
and the magnetization in the bulk of both ferromagnetic regions points along $\hx$. For a given $n$, the helicity $p$ and the function $\varphi_0$ are determined by the interface conditions. In the composite magnet consider here, where $q_u(L-L_0)=200$ is very large, the values of $p$ are given by the solutions to \cite{Laliena2024}
\be
p - 1 = -\frac{\rho q_u}{q_c} \, \sin(pq_cL_0+n\pi), \label{eq:p}
\ee
and we have
\be
\varphi_0(z) = -2\atan\e^{-q_u(z-z_0)}, \label{eq:phi0}
\ee
where $z_0$ is given by
\be
\tanh\big(q_u(L_0-z_0)\big) = \cos(pq_cL_0+n\pi). \label{eq:z0}
\ee
Thus, $\varphi_0$ has the form of a standard domain wall centered at $z_0$. Only states with $z_0<L_0$ are stable (this constitutes half of the solutions of Eq. (\ref{eq:p}), see Ref. \onlinecite{Laliena2024}).  That is, the stable states have the center of the domain wall in the non-physical region of the uniaxial ferromagnet. States with the center of the wall in the physical region ($L_0<z<L$) are so strongly penalized energetically that they become unstable. In addition, the states are stable only if $\pmin<p<\pmax$, with $\pmin=1-\sqrt{\hcrit}$ and  $\pmax=1+\sqrt{\hcrit}$, where $\hcrit=1-\kappa_c=6.17$ is the dimensionless critical parallel field of the monoaxial chiral magnet \cite{Laliena2024}. We call $p$ the \textit{helicity} of the state, and the solutions of Eq. \eqref{eq:p} for a given $n$ that yield stable helical states are referred to as the \textit{admissible values} of helicity of type $n$.

\begin{figure}[t!]
\centering
\includegraphics[width=0.45\textwidth]{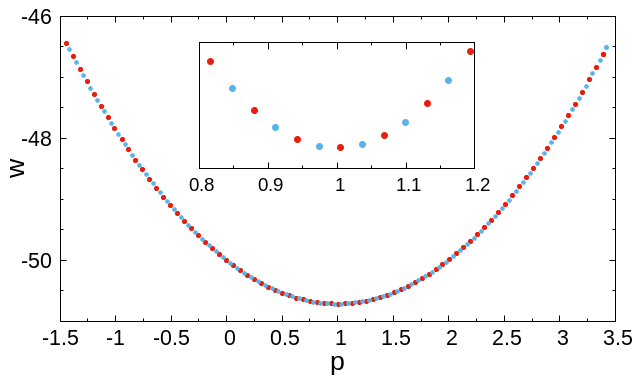}
\captionof{figure}{Energy density of the helical states, in units of $A_cq_c^2$, as a function of $p$. The red and blue dots correspond to $n=0$ and $n=1$, respectively.
 \label{fig:energy}}
\end{figure}

The admissible values of $p$ are almost uniformly distributed in $[\pmin,\pmax]$, alternating in type: the two closest states (in terms of $p$) to a state of type 0 are of type 1, and vice versa, and their number increases linearly with the size of the chiral magnet along the chiral axis, $L_0$. The energy of the stable helical states as a function of $p$ is shown in Fig. \ref{fig:energy}. The absolute minimum of the energy is attained at $p \approx 1.005$ and is of type 0. Despite appearances, the other states are metastable because the value of $p$ cannot be changed continuously due to the dynamical topological protection discussed in Sec. \ref{sec:topo}.

With the help of the dynamic code we can verify that the helical states with admissible values of helicity are actually metastable, and that they are the only metastable helical states \footnote{The analysis of stability of Ref. \cite{Laliena2024} was limited to linear stability, which is a necessary but not sufficient condition for stability. We are usually forced to rely on linear stability, since the complete stability analysis is overly complicated.}. To this end, we prepare an initial state of the form (\ref{eq:phi}) with a non-admisible value of $p$ for a chosen $n$. In the ferromagnetic regions, the magnetization is uniform and points in the $\hx$ direction, $\varphi_0(z)=1$. The magnetization is thus discontinuous at the interfaces, and thus this vector field does not belong to the set of possible magnetization states. However, in the discretized problem, the state can be considered ``continuous'', with a very sharp variation of $\hm$ across the interfaces, and the discontinuity posses no practical problem. 

The initial state is allowed to evolve until an equilibrium state is reached. The resulting equilibrium state is a helical state with an admissible $p$ close to the initial, non-admissible $p$. The system does not evolve toward the absolute minimum fo the energy but instead gets trapped in a helical state of admissible helicity close to the initial $p$. The same results are obtained if random noise is added to the initial state. The final state is therefore metastable, in agreement with the linear stability analysis of Ref. \onlinecite{Laliena2024}.

\begin{figure}[t!]
\centering
\includegraphics[width=0.23\textwidth]{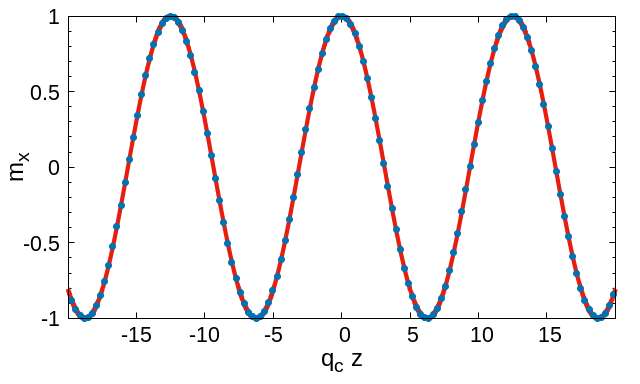}
\includegraphics[width=0.23\textwidth]{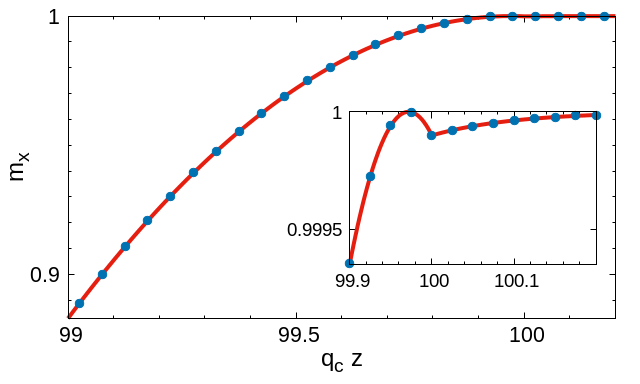}

\includegraphics[width=0.23\textwidth]{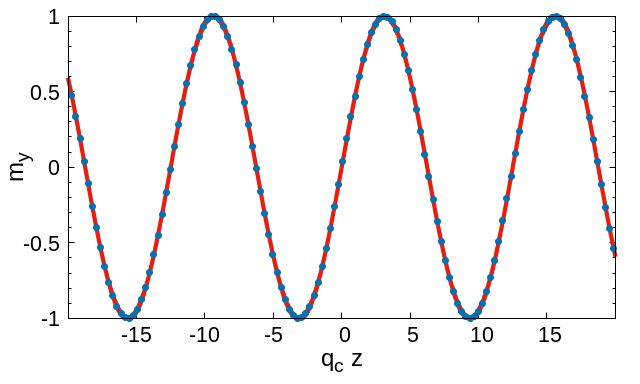}
\includegraphics[width=0.23\textwidth]{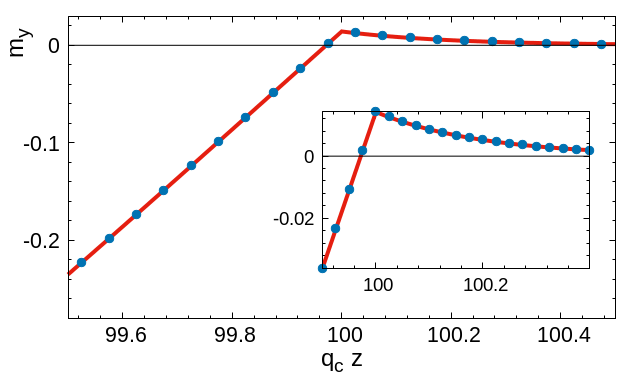}

\includegraphics[width=0.4\textwidth]{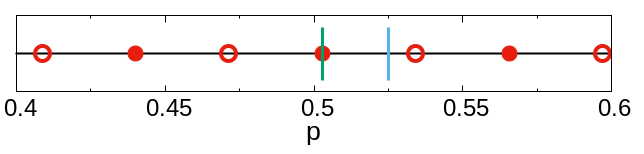}

\captionof{figure}{Magnetization components of the metastable helical state reached after allowing an initial state with $n=0$ and a non-admissible helicity of $p=0.525$ to evolve. The left panels display the center of the chiral magnet, and the right panels show the vicinity of one of the interfaces (the insets provide a zoomed-in view of the interface). The blue dots represent the numerical results, and the red line corresponds to the exact helical state with $p=0.503$. In the bottom panel, the admissible values of $p$ near the initial non-admissible value are shown as filled ($n=0$) and open ($n=1$) red circles. The initial value of $p$ is indicated by the blue vertical line segment, and the helicity of the final metastable state by the green segment.
 \label{fig:exact}}
\end{figure}

Figure \ref{fig:exact} shows the magnetization of the metastable state reached when the initial helical state has a non-admissible $p=0.525$ and $n=0$. The metastable state has $n=0$ and the admissible helicity $p=0.503$. The exact helical state given by Eqs. \eqref{eq:phi}--\eqref{eq:z0} is also shown. An excellent agreement is observed between the numerical and exact solutions. This metastable state has also been calculated numerically using the static code. The results obtained with the static and dynamic codes cannot be distinguished on the scale of Fig. \ref{fig:exact}. Therefore, a check of both numerical codes is obtained as a byproduct of these results.

\section{Dynamical topological protection\label{sec:topo}}

What we call helicity of a helical state coincides almost exactly with the winding number density, \textit{i.e.}, the number of times that the magnetization turns around the chiral axis, divided by the length of the chiral magnet along the chiral axis. For a any state with magnetization $\vm$ modulated only along $z$, we define the winding number density by the formula
\be
p = \frac{1}{2q_cL_0}\int_{-L}^{L} \frac{\hz\cdot(\vm\times\vm^{\,\prime})}{|\hz\times\vm|^2}\,dz, \label{eq:pdef}
\ee
which is meaningful as far as $|\hz\times\vm|>\delta$ for some $\delta>0$.
In Eq. (\ref{eq:pdef}) we divide by $2q_cL_0$ instead of by the total length $2q_cL$, which at first sight may seem more natural. We do so in order to make winding number density (almost) coincident with helicity for helical states:
it is straightforward to check that, with the definition of Eq. (\ref{eq:pdef}), the difference between winding number density and helicity is very small. That is why, to avoid symbol proliferation, we use the same symbol $p$ for both quantities. Since in the helical states and in the states that are dynamically reached by applying the external stimuli considered in this work only a fraction of a magnetization turn takes place within the ferromagnetic slabs, the winding number density would be very small with the apparently natural definition if $L-L_0$ were very large. In other words, since winding number is essentially supported within the chiral magnet, Eq. (\ref{eq:pdef}) is actually the natural definition of winding number density. For future use, we call the integrand in the right-hand side of Eq. (\ref{eq:pdef}) the \textit{local winding number density}.

\begin{figure}[t!]
\centering
\includegraphics[width=0.23\textwidth]{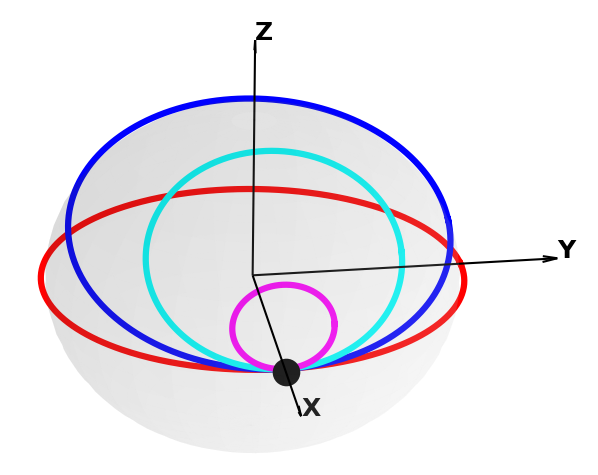}
\includegraphics[width=0.22\textwidth]{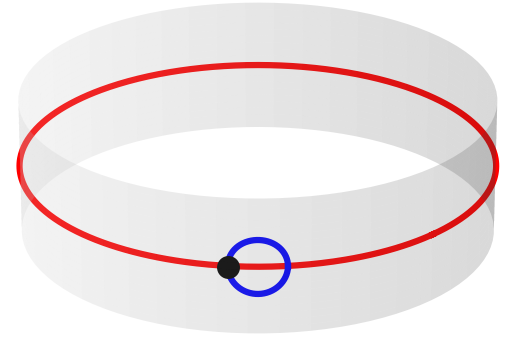}
\caption{Illustration of the homotopy of curves on a sphere and a cylinder. On the sphere (left panel), to transform continuously a closed curve of winding number one (red) into a closed curve of zero winding number (purple), or vice versa, one of the poles must be crossed. On the cylinder (right panel), such a continuous transformation is impossible.
 \label{fig:homotopy}}
\end{figure}

In the states that appear in this work, the magnetization inside the ferromagnetic slabs is nearly uniform, except in the vicinity of the internal interfaces, and points out in the same direction in both slabs. Therefore, if we plot on the unit sphere the magnetization fields of  any of these, we obtain a closed curve, parametrized by the coordinate $z\in[-L,L]$ (Fig. \ref{fig:homotopy}, left). This implies that the \textit{winding number} $2q_cL_0 p$, which is defined for curves that do not intersect the poles ($\hz$ is the polar axis), is an \textit{integer} that counts the number of times that the magnetization turns around the chiral axis $\hz$, and the terms winding number and winding number density are justified.

Helical states can be classified by an integer, the winding number, instead of a real number, the helicity. We prefer helicity (and winding number density, for more general states) to winding number , since this is a large number that scales with $L_0$, while $p$ is a real number contained in $[\pmin,\pmax]$ and nearly independent of $L_0$. For instance, the equilibrium state has winding number $201$ and $401$ for $L_0=200$ and $L_0=400$, while $p=1.005$ in the former case and $p=1.0025$ in the latter.

Winding number is not a topological quantity, even if the magnetization curves on the unit sphere are closed, since one closed curve can be continuously transformed to another one with different winding number. Hence, winding number can be created or destroyed without any discontinuity in the magnetization field. In mathematical jargon, the first homotopy group of the sphere is trivial. We notice, however, that to change the winding number the magnetization curve has to cross one of the poles of the sphere (see Fig. \ref{fig:homotopy}, left). The magnetization is continuous but the winding number changes discontinuously (by an integer) due to the singularity of the integrand of Eq. (\ref{eq:pdef}).
 
The magnetic anisotropy and the DMI of the chiral magnet make energetically extremely unfavorable magnetization fields whose curves on the unit sphere are far from the equator. This means that, effectively and for energetic reasons, the magnetization takes values on a certain cylinder centered around the equator, and excursions of the magnetization to the poles are severely suppressed. Winding number on a cylinder is a topological quantity: a closed curve on a cylinder cannot be continuously deformed to another curve with different winding number (Fig. \ref{fig:homotopy}, right). The first homotopy group of the cylinder is non trivial. Winding number cannot be created or destroyed without discontinuities in the magnetization, which are micromagnetically forbidden. Winding number can be changed only by large external stimuly (applied field and currents) that inject into the system enough energy to cross the poles.

We term the fact that the magnetization is confined to a cylindric subset of the unit sphere by energy barriers the \textit{dynamical topological protection} of the magnetization. Thanks to this effect, winding number becomes effectively a topological quantity. This explains the stability of the helical states discussed in the previous section and many of the results presented in sections \ref{sec:conical} and \ref{sec:static}. For instance, that winding number density is preserved under the application of magnetic fields and currents of intensity below some threshold, and is well defined and constant even for time dependent magnetic configurations.

The topological protection requires that the magnetization field describe only closed curves on a cylindric subset of the unit sphere. It is clear that winding number can be created or destroyed if open curves are allowed: any open curve on a cylinder, whatever its winding number, can be continuously schrunk to a point. In a chiral magnet in contact only with non magnetic media the boundary conditions do not force the magnetization  to have the same values at the boundaries. This means the magnetization curves on the unit sphere are generally open, and winding number can be easily created or destroyed. Dynamical topological protection is not effective, and any helical state relaxes, decreasing its energy, to the equilibrium state with $p=1$. That is why metastable helical states have not been detected in monoaxial chiral magnets.

The discussion of the above paragraph elucidates completely the role of the ferromagnetic slabs in stabilizing helical states other than the equilibrium state. Despite the Neumann boundary conditions, the nature of the ferromagnet fixes the value of the magnetization at the boundaries, thereby ensuring that the magnetization curves on the unit sphere are closed. Opening the curves, by reversing the magnetization on the slabs, requires a big amount of energy that can be provided only by electric currents of rather high strength. 

In summary, the dynamical topological protection that stabilizes numerous helicoidal states rests on two pillars: first, the presence of the ferromagnetic slabs, which forces the magnetization curves on the unit sphere to close, and second, the combination of uniaxial DMI and high anisotropy in the chiral magnet, which effectively restricts these curves to a cylindrical equatorial band.

\begin{figure}[t!]
\centering
\includegraphics[width=0.23\textwidth]{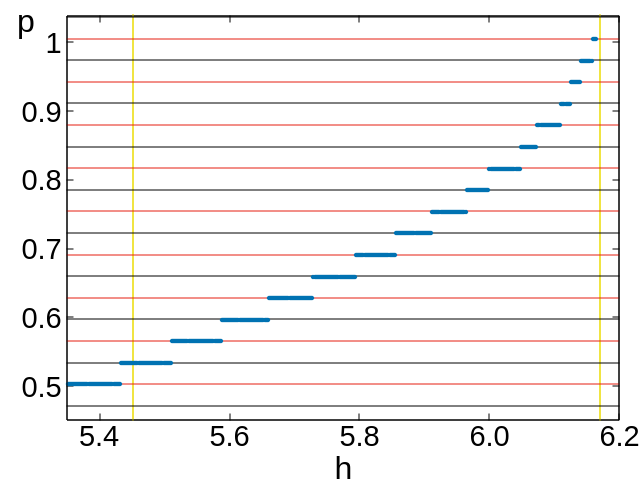}
\includegraphics[width=0.23\textwidth]{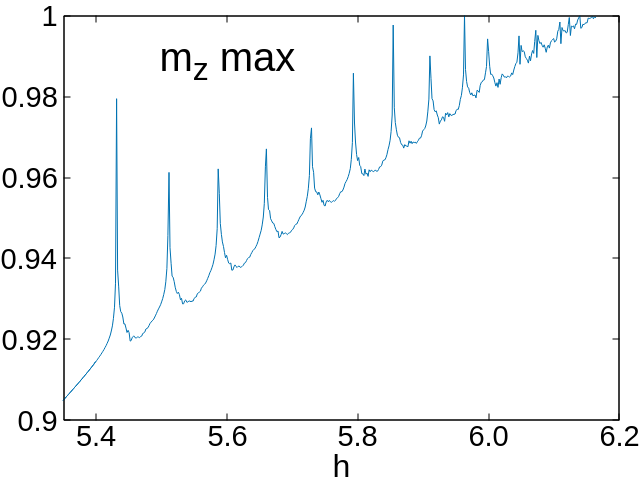}
\captionof{figure}{(Left) Evolution of the helicity as a magnetic field is slowly applied to the helical state with $p=0.503$, starting at $h=0$ and ending at the critical field $\hcrit=6.17$. The abscissa represents the instantaneous applied field. The horizontal lines indicate the admissible values of $p$ for $n=0$ (red) and $n=1$ (black). The vertical yellow lines indicate the destabilizing field for $p=0.503$ and the critical field $\hcrit=6,17$. (Right) Maximum value of $m_z$ within the chiral magnet as a function of the instantaneous magnetic field. The sharp, narrow peaks coincide with the jumps in helicity.
 \label{fig:process_v0}}
\end{figure}

\section{Conical states under an applied field\label{sec:conical}}

The helical states survive the adiabatic (slow) application of a magnetic field directed along the chiral axis, $\hz$. The magnetization acquires a component along $\hz$ and resembles a conical state within the chiral magnet region, slightly distorted in the vicinity of the interfaces to fulfill the interface conditions. The helicity, $p$, is preserved under the applied field, provided it is not too intense, due to the dynamical topological protection. If the applied field strength exceeds a certain threshold, the helicity jumps to another close admissible value. We call this threshold the \textit{destabilizing field} of the state.

The behavior of the helicity when a magnetic field is applied slowly to the helical state with $p=0.503$ ($n=0$), until the critical field ($\hcrit=6.17$) is reached, is shown in Fig. \ref{fig:process_v0} (left). Instead of time, the abscissa represents the instantaneous applied field. The helicity remains constant until $h\approx 5.45$, at which point it jumps to the helical state with the next higher admissible helicity, $p=0.535$ ($n=1$). As the field increases further, the helicity continues to increase via discrete jumps to other admissible values until, in the vicinity of $\hcrit$ (indicated by the yellow vertical line on the right side of the figure), it reaches $p=1.005$ ($n=0$), which corresponds to the absolute minimum of the energy. 
If, instead, the initial state has an admissible $p>1.005$, the helicity jumps to lower admissible values as the field increases, eventually reaching the state with $p=1.005$ near $\hcrit$. The effect of applying a magnetic field is thus to drive the helicity toward the energy minimum, $p=1.005$. It is a means to create helicity if the initial state has $p<1.005$, and to destroy it if $p>1.005$.

\begin{figure}[t!]
\centering
\includegraphics[width=0.35\textwidth]{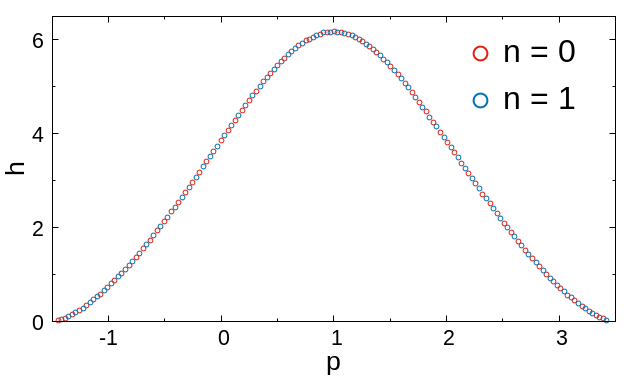}
\captionof{figure}{ Destabilizing field for helical states as a function of $p$. The red and blue dots correspond to $n=0$ and $n=1$, respectively. The destabilizing field for the equilibrium state ($p=1.005$) coincides with the critical field $\hcrit=6.17$, as it must be.
 \label{fig:critical_h}}
\end{figure}

The change of helicity is only possible if the magnetization component along the chiral axis, $m_z$, reaches its maximum value, $m_z=1$, at some point within the chiral magnet (see Sec. \ref{sec:topo}). Figure \ref{fig:process_v0} (right) shows the maximum of $m_z$ as a function of the instantaneous applied field. The sharp, narrow peaks coincide with the jumps in helicity. The reason why the maximum of $m_z$ does not appear to attain the value of one in Fig. \ref{fig:process_v0} (right) is a lack of time resolution.

We conclude that each conical state is characterized by an admissible value of the helicity, and has a destabilizing field above which the state becomes unstable. The destabilizing field can be computed with the static code and is displayed as a function of $p$ in Fig. \ref{fig:critical_h}. Note that all data points, irrespective of the value of $n$, lie on a smooth curve. The strength of the destabilizing field is the same whether it is applied along the positive or negative direction of the chiral axis. The vertical yellow line on the left side of Fig. \ref{fig:process_v0} (left) marks the destabilizing field for $p=0.503$ computed with the static code. It coincides almost perfectly with the instantaneous field at which the helicity jumps from $p=0.503$ to $p=0.535$. The destabilizing field for each admissible $p$ has also been calculated using a relaxation version of the dynamic code. On the scale of Fig. \ref{fig:critical_h}, the results obtained with the dynamic and static codes cannot be distinguished.

\begin{figure}[t!]
\centering
\includegraphics[width=0.4\textwidth]{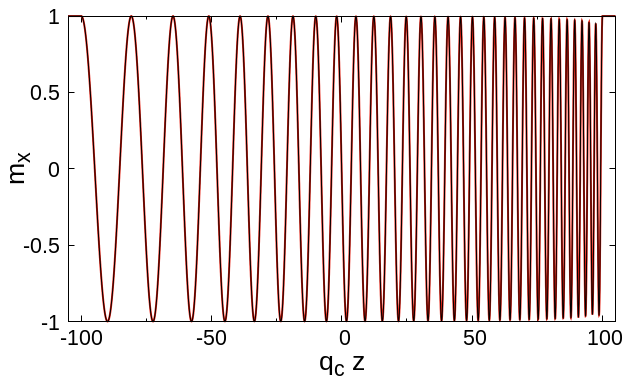}
\includegraphics[width=0.4\textwidth]{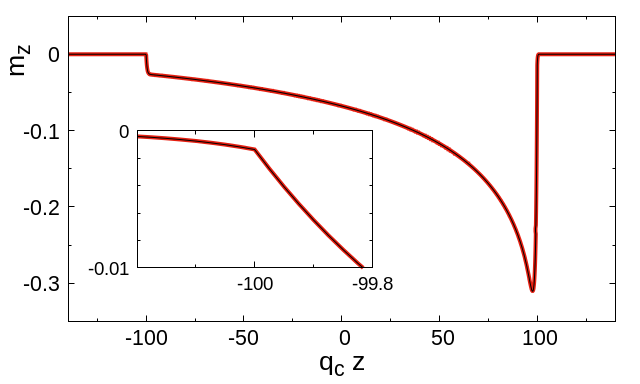}
\captionof{figure}{Magnetization components $m_x$ (top) and $m_z$ (bottom) of the static state obtained by applying a current $v=-0.5$ to the helical state with $p=1.005$. The final static state (shown here) has the same $p$, but the winding number density is inhomogeneously distributed, concentrating in the vicinity of $q_c L_0=100$.
 \label{fig:static}}
\end{figure}

\section{Response to a spin-polarized electric current \label{sec:static}}

If a polarized current is applied to a helical state of chiral magnet without boundaries, the magnetization reaches a steady motion state, moving rigidly with constant velocity against the current. The effect of the current in the limited magnet considered in this work is far from obvious, since the steady motion state is clearly impossible, or at least difficult to conceive, as it should take place in the ferromagnetic slabs in the form domain wall motion. What actually happens is that the steady motion state is replaced by a static state, with a time independent magnetization. The winding number density is kept constant, again due to the dynamical topological protection, so that the static state has the same value of $p$ than the initial helical state. Fig. \ref{fig:static} displays the static state attained after the slow application to the state with $p=1.005$ ($n=0$) of a current that reaches a final value $v=-0.5$. The red lines corresponds to the state reached by the time evolution, computed with the dynamic code; the black lines, to the static state computed with the static code for $h=0$ and $v=-0.5$. The aggreement is excellent (the lines cannot be distinguished in the scale of the figures).

The effect of the current is twofold. First, it induces an inhomogeneous magnetization component along the chiral axis, whose absolute value increases against the current (bottom panel of Fig. \ref{fig:static}). Second, the local winding number density within the chiral magnet becomes inhomogeneous, concentrating in the vicinity of the interface located upstream with respect to the current (top panel of Fig. \ref{fig:static}). Winding number (and thus the average winding number density), however, is preserved; the turns of the magnetization around the chiral axis pile up near the interface, but the number of turns remains constant due to dynamical topological protection. If the current is switched off, the initial metastable helical state is recovered.

\begin{figure}[t!]
\centering
\includegraphics[width=0.23\textwidth]{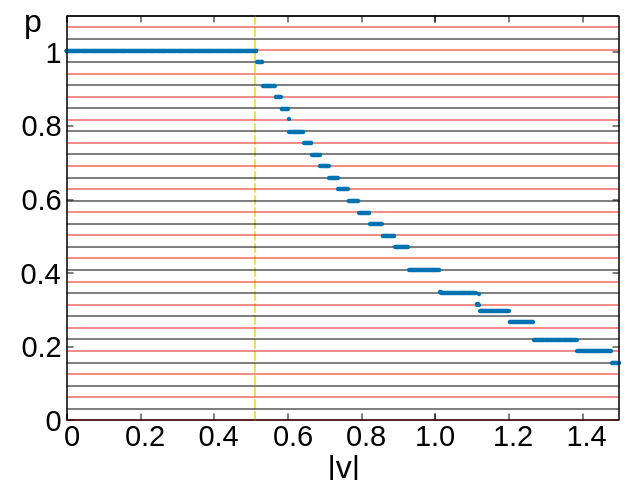}
\includegraphics[width=0.23\textwidth]{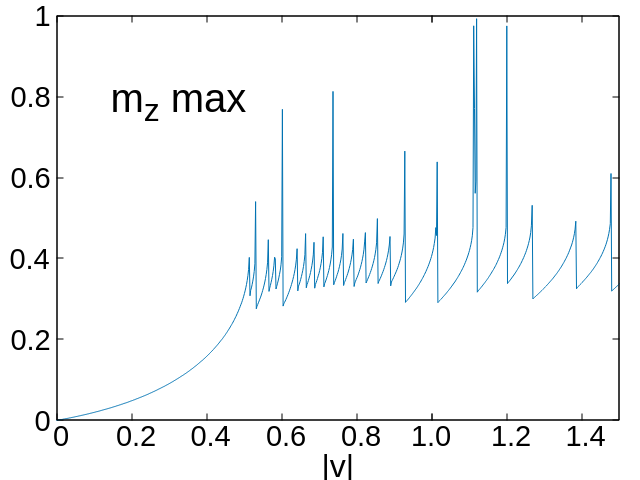}
\captionof{figure}{Evolution of the helicity when a current in the $-\hz$ direction is slowly applied to the helical state with $p=1.005$. The final current is $v=-1.5$. The horizontal lines indicate the admissible values of $p$ for $n=0$ (red) and $n=1$ (black).
 \label{fig:process_h0}}
\end{figure}

The dynamic process leading to the static state proceeds as follows. Initially, the magnetization is set in motion by the current, just as it happens in the magnet without boundaries. The winding number density, however, is unable to penetrate into the ferromagnetic slabs and accumulates in the vicinity of the interface. It is tempting to interpret this fact pictorially by supposing that the magnetization is set in motion by a pressure exerted by the current on the winding number (or "helical turns"), and that the ferromagnet exerts an opposite pressure to prevent the winding number density from penetrating into it. The static state is reached when both pressures balance.

By increasing the current (along the negative $\hz$ axis), the static state with $p=1.005$ becomes unstable, and the system evolves toward another static state, with the value of $p$ jumping to a lower admissible value. Fig. \ref{fig:process_h0} (left) displays this behavior, which is the same regardless of the initial helical state. It is remarkable that the values of $p$ remain within the admissible set for any static state, although the static states in the presence of a current are very different from the helical states at zero current (Fig. \ref{fig:static}). This observation seems to contradict Fig. \ref{fig:process_h0} (left), where it is observed that when $\vert{}v\vert{}$ is higher than about $1.1$, the helicity jumps to values that lie halfway between two admissible values. What happens in these cases is that, besides the jump in helicity, the magnetization in the ferromagnetic slab located upstream with respect to the current flow is reversed. For such large currents, the "pressure" exerted by the ferromagnet is not enough to hold back the winding number, which enters the ferromagnet in the form of domain walls, leading to magnetization reversal in the ferromagnetic slab. A static state is reached in which the magnetization has opposite directions in each ferromagnetic slab. The winding number density of these static states corresponds to one of the admissible values of the helicity of antisymmetric helical states, with $n=1/2$ or $n=3/2$, which have opposite magnetizations in each ferromagnetic slab, and whose symmetry is $\hx\cdot\vm(-z)=-\hx\cdot\vm(z)$, $\hy\cdot\vm(-z)=\hy\cdot\vm(z)$. These states are also interesting but are not considered in this work, since we want to keep the magnetization conditions constant in the ferromagnetic slabs. To avoid these, current densities higher than $\vert{}v\vert{}\approx 1$ have to be avoided.

Winding number density can only be changed if the magnetization component along the chiral axis, $m_z$, reaches its maximum value, $m_z=1$, at some point. By comparing the left and right panels of Fig. \ref{fig:process_h0}, it is observed that the jumps in $p$ coincide with the sharp peaks in the maximum value of $m_z$, which is not seen to attain the value one due to the lack of temporal resolution.

As it happens with applied magnetic fields, each helical state has two destabilizing currents: one positive, $v_+ > 0$, and one negative, $v_- < 0$. If, at zero field, a current $v\in(v_- ,v_+)$ is applied to a helical state with helicity $p$, the system reaches a static state characterized by an inhomogeneous local winding number density, while preserving the initial winding number density, $p$. If the current is switched off, the original helical state is recovered. We may thus say that the helical state is stable under the application of currents in the interval $(v_-, v_+)$. If the applied current, however, lies outside this interval, the value of $p$ jumps to a lower admissible value, and the system reaches a new static state with this reduced winding number density. If the current is then switched off, the helical state corresponding to this lower helicity is obtained. Applying an electric current at zero field is therefore a way to decrease the helicity of a state.

\begin{figure}[t!]
\centering
\includegraphics[width=0.4\textwidth]{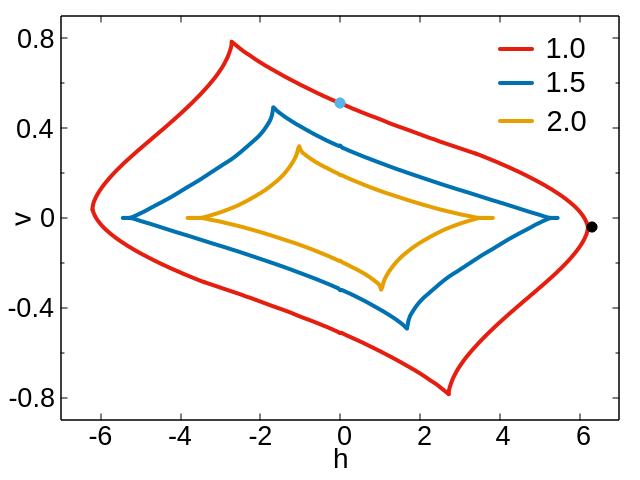}

\includegraphics[width=0.4\textwidth]{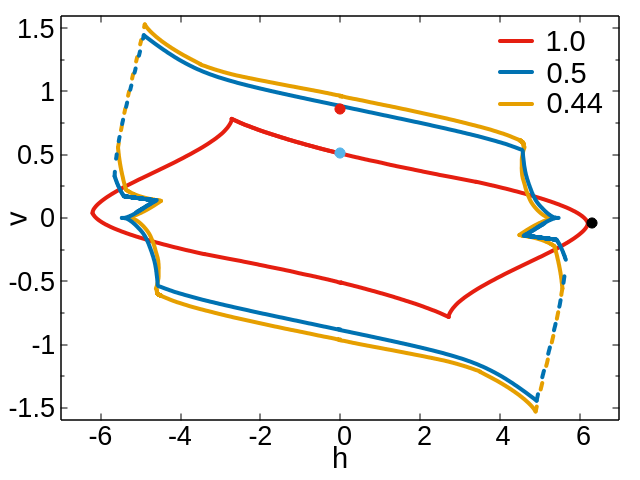}
\captionof{figure}{Domain of stability of helical states with $p\geq 1$ (top) and $p\leq 1$ (bottom) for the values of $p$ shown in the legend (all for $n=0$).
 \label{fig:phd}}
\end{figure}

The simultaneous application of a magnetic field and an electric current to a helical state also leads to a static state that preserves the winding number density, $p$, provided the field and current intensities are not too large. If these intensities exceed certain threshold values, a static state with a different admissible $p$ is reached. Thus, to each admissible $p$ there corresponds a region in the $(h,v)$ plane where a static state with winding number density $p$ exists. We call this region the \textit{stability domain} of $p$; outside of it, no static state with winding number density $p$ can exist.

The stability domains for several values of $p$ are displayed in Fig. \ref{fig:phd}. In the top panel, we show results for $p \geq 1$. We see that static states with $p > 1$ cannot be reached from states with smaller $p$ through the quasistatic application of currents and fields. Indeed, if $p > 1$, the quasistatic application of a field and current always leads to a decrease in $p$. The bottom panel displays the stability domains for values $p \leq 1$, where the situation is different. For instance, starting with $p = 1.005$ at $h = 0$ and $v = 0$, and slowly applying a field and current pulse, we can reach a point in the $(h,v)$ plane that lies outside the stability domain of $p = 1.005$ (bounded by the red line in Fig. \ref{fig:phd}) but inside the stability domain of $p = 0.503$ (bounded by the blue line). If it is sufficiently close to the boundary of the stability domain of $p = 0.503$,  as the red point of Fig. \ref{fig:phd} (bottom), the system will relax into a static state with $p = 0.503$, since any static state with a higher $p$ is unstable there. By switching off the pulse, we will end up with the helical state with $p = 0.503$ at $h = 0$ and $v = 0$. If we want to restore the state with $p = 1.005$, we can apply a field and current pulse that brings the system outside the stability domain of $p = 0.503$ and reaches a point close to the boundary of the stability domain of $p = 1.005$, as the black point of Fig. \ref{fig:phd} (bottom). In this way, we reach a static state with $p = 1.005$, and by removing the field and current, we obtain the helical state with $p = 1.005$ at zero field and current. The dynamical feasibility of these mechanisms is studied in the next section.

Looking at the bottom panel of Fig. \ref{fig:phd}, the reader will notice that some portions of the stability boundaries for states with $p<1.0$ are plotted as dashed lines. These portions are located in regions characterized by high magnetic fields and large currents directed against the field. Under such conditions, the energy barriers between static states are very low, and the system easily jumps from one state to another. This makes it difficult to pinpoint the exact boundaries accurately. Nevertheless, it is quite clear that, in these regions, all static states are destabilized at roughly the same points, represented by the dashed lines in the bottom panel of Fig. \ref{fig:phd}. This means that these regions are unsuitable for switching between helical states, as the final state would be uncertain.

\section{Switching between helical states \label{sec:switch}}

The mechanism to switch between states with different helicities proposed in the previous section has been verified by numerical computations using the dynamic code. We illustrate it by analyzing two processes in detail: in the first, the aim is to switch to the metastable helical state with $p=0.503$ ($n=0$), starting from the global equilibrium state with $p=1.005$ ($n=0$); in the second process, the initial and final states are interchanged.

\begin{figure}[t!]
\centering
\includegraphics[width=0.23\textwidth]{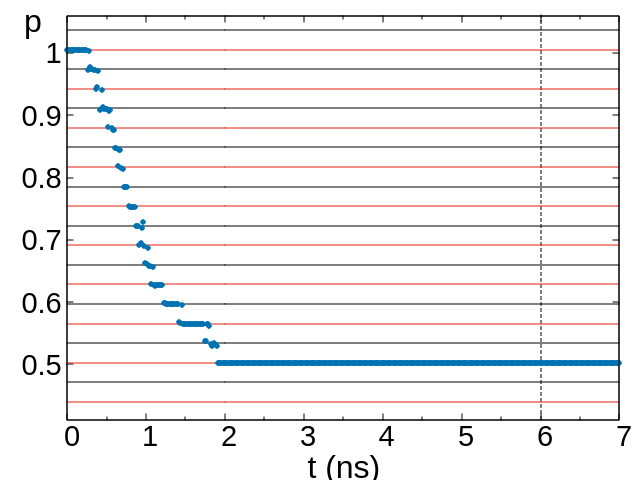}
\includegraphics[width=0.23\textwidth]{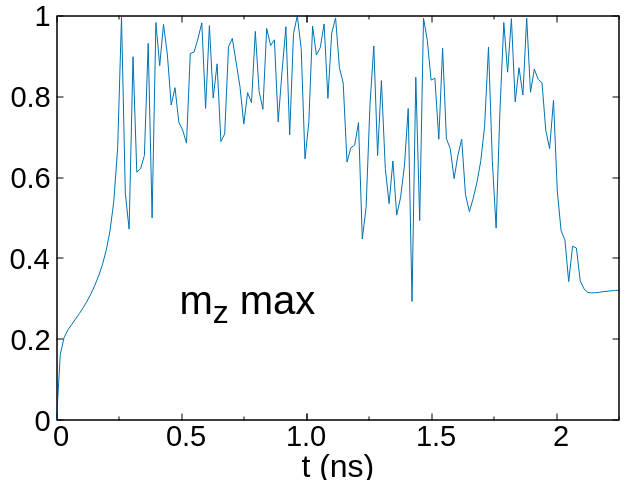}

\includegraphics[width=0.23\textwidth]{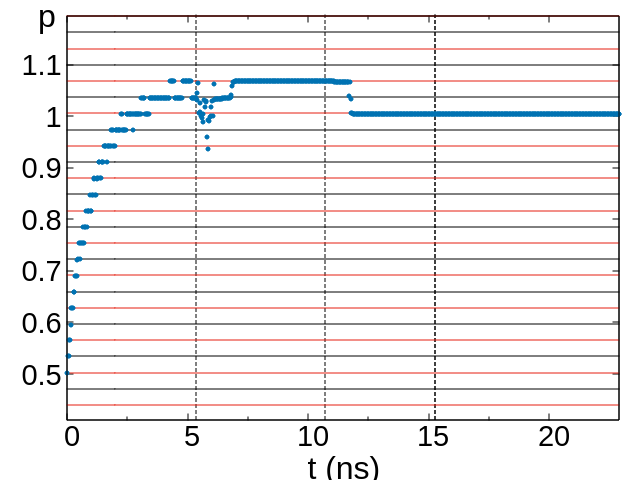}
\includegraphics[width=0.23\textwidth]{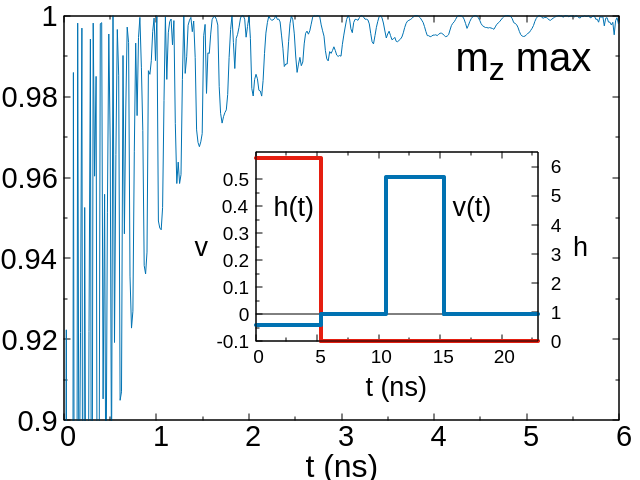}
\captionof{figure}{Switching between helical states by applying field and current pulses. Top panels: the initial state has $p=1.005$ ($n=0$) and the final state has $p=0.503$ ($n=0$). Bottom panels: the initial state has $p=0.503$ ($n=0$) and the final state has $p=1.005$ ($n=0$). The vertical lines in the left panels mark the instants at which the pulses are switched on or off.
 \label{fig:switch}}
\end{figure}

Starting from $p=1.005$ at $h=0$ and $v=0$, we abruptly apply a square current pulse of intensity $v=0.86$ and a duration of 6 ns, after which the system is allowed to relax, at $h=0$ and $v=0$, for another 12 ns. The current pulse brings the external parameters $h$ and $v$ to the red point in Fig. \ref{fig:phd} (bottom). The evolution of the winding number density is displayed in the top left panel of Fig. \ref{fig:switch}. The current destroys winding number in the first 2 ns, after which the system is left in the static state with $p=0.503$, which is the highest stable $p$ at $h=0$ and $v=0.86$, as seen in Fig. \ref{fig:phd} (bottom). The value of $p$ remains unchanged when the current is removed at $t=6$ ns, and after relaxation we get the helical state with $p=0.503$ at $h=0$ and $v=0$, which was our goal. The top right panel of Fig. \ref{fig:switch} shows how the jumps in helicity during the first 2 ns are accompanied by sharp peaks in the maximum of $m_z$, which reaches the limit value of $m_z=1$.

The process of replacing $p=0.503$ with $p=1.005$ is somewhat harder because creating helicity is more difficult than destroying it. To this end, we apply combined square field and current pulses to the helical state with $p=0.503$, initially at $h=0$ and $v=0$. The whole process consists of four stages. In the first stage, a square pulse of field and current with intensities $h=6.3$ and $v=-0.04$, and a duration of 5.3 ns, is applied. These external stimuli bring the system to the point in the $(h,v)$ plane marked by the black dots in Fig. \ref{fig:phd} (top and bottom). In the second stage, the field and the current are removed and the system is allowed to relax at $h=0$ and $v=0$ for another 5.3 ns. In the third stage, a square current pulse of intensity $v=0.51$ and duration 4.7 ns is applied. Again, this external stimulus defines a point in the $(h,v)$ plane, marked by the blue dots in Fig. \ref{fig:phd} (top and bottom). Finally, in the fourth stage, the current is removed and the system is allowed to relax at $h=0$ and $v=0$ for 7.7 ns. The time dependence of the external parameters for the entire process, which has a total duration of 23 ns, is graphically displayed in the inset of the bottom right panel of Fig. \ref{fig:switch}.

The evolution of the winding number density during the second process is displayed in the bottom left panel of Fig. \ref{fig:switch}. The value of $p$ increases almost monotonically to values higher than 1.0 during the first stage, since all static states with $p<1$ are destabilized by the combination of field and current (Fig. \ref{fig:phd}, bottom). The bottom right panel of Fig. \ref{fig:switch} shows the sharp peaks in the maximum value of $m_z$ that accompany the helicity jumps.At $h=6.3$ and $v=-0.04$ (black dots in Fig. \ref{fig:phd}), there is no static state, and the magnetization evolves over time in an almost periodic manner, keeping a constant value of $p$ higher than unity. In the second stage, the field and current are switched off, and the magnetization relaxes to a time-independent metastable helical state. During the transient, the value of $p$ oscillates around one and is finally stabilized at $p=1.067$ ($n=0$), which is the allowed value higher than $1.005$ and closest to the target value $p=1.005$. The third stage of the process is designed to bring any possible state with $p>1.005$ reached in the second stage ($p=1.067$ in the present case) to the state with $p=1.005$ by applying a current pulse of intensity $v=0.15$ at $h=0$. This pulse brings the external parameters $h$ and $v$ to the blue point in Fig. \ref{fig:phd} (bottom). At this point, all states with $p>1.005$ are unstable, and the winding number density jumps to $p=1.005$ and reaches the corresponding static state. The final state, stabilized after switching off the current (fourth stage), is the equilibrium helical state with $p=1.005$.

\section{Discussion and conclusions \label{sec:conc}}

A monoaxial chiral magnet attached two slabs of a uniaxial ferromagnet in the way described in \ref{sec:model} (see Fig. \ref{fig:composite_magnet}) has a multiplicity of metastable helical states that are distinguished by the value of their winding number density, which coincides with helicity and is denoted by $p$. The values $p$ of the helicity of metastable states form a large finite set which we call the \textit{admissible values} of helicity. These metastable states are stabilized thanks to the dynamical topological protection discussed in section \ref{sec:topo}, which is absent if the monoaxial chiral magnet is in contact only with non-magnetic media. This fact explains why these metastable states have nerver been observed.

Dynamical topological protection is expected to render metastable helical states long-lived. Indeed, numerical simulations show how difficult it is to change helicity. Thus, the helical states may be useful in spintronics and magnonics, as their magnetic properties depend on $p$. Clearly, their utility would be greatly enhanced by the ability to switch between them. In this work, we have studied the static and dynamic properties of the helical states and conclude that switching can be achieved by means of applied magnetic fields and spin-polarized electric currents.

In summary, owing again to the dynamical topological protection, the helicity $p$ is preserved under a magnetic field applied along the chiral axis, provided the field intensity is not excessively high. If the intensity is large enough, the state with a given $p$ becomes unstable and is replaced by a different state with another admissible value of $p$. This instability occurs when the magnetic state acquires sufficient energy to overcome the barrier set by the magnetic anisotropy and DMI of the chiral magnet, thereby rendering the dynamical topological protection ineffective.

The effect of an applied current on a helical state is, at first sight, more intriguing. In a magnet without boundaries, the magnetization reaches a steady-state motion, moving rigidly with constant velocity in the direction opposite to the current. This is clearly impossible in the bounded magnet considered in this work. In this case, the steady motion is replaced by a static state with a time-independent magnetization. The winding number density is kept constant, once again by virtue of the dynamical topological protection, and this static state retains the same value of $p$ as the initial helical stat 
The local winding number density of the static state is not homogeneous, though; rather, it is concentrated near the interface where the current flows from the ferromagnet into the chiral magnet. In other words, the "helical turns" are piled up in the vicinity of this interface. Initially, the magnetization is set in motion by the current, as occurs in the magnet without boundaries. However, the ferromagnet repels the winding number density so that, instead of being homogeneously distributed—as in the helical state at zero current—it accumulates near the interface. It is tempting to interpret pictorially this result by supposing that the magnetization is set in motion by a pressure exerted by the current on the winding number density (or "helical turns"), and that the ferromagnet exerts an opposing pressure to prevent the winding number from penetrating it. The static state is reached when both pressures are balanced.

The dynamical topological protection of a static state is lost if the field or current intensities are large enough to align the magnetization with the chiral axis at some point. In this case, the static state is replaced by another one with a different, yet close, $p$ belonging to the set of admissible values.Thus, each static state, characterized by an admissible $p$, possesses a stability domain in the parameter plane defined by the strengths of the magnetic field and the current. The determination of these stability domains, carried out in this work, is crucial for designing mechanisms to switch between different helical states. We show (Section \ref{sec:switch}) that switching between two helical states, with initial and final helicities $p_i$ and $p_f$, is feasible by applying appropriate combined pulses of field and current. These pulses bring the external parameters to a region in the parameter plane where the static state with winding number density $p_i$ becomes unstable and is replaced by the state with winding number density $p_f$. Once the field and current are switched off, the relaxed state retains the helicity $p_f$.

Although this work focuses on monoaxial chiral magnets, it is rather clear that dynamical topological protection also holds in cubic chiral magnets attached to ferromagnetic slabs. However, in materials such as MnSi, the low magnetic anisotropy would presumably result in much lower destabilizing fields and currents than in monoaxial chiral magnets (which may be an advantage rather than a disadvantage). Composite chiral cubic magnets systems should therefore host metastable helical states with properties similar to those studied here.

There is a large body of experimental research on helical states of chiral magnets, but all experiments to date have been carried out on magnets in contact with air or non-magnetic media. Under these conditions, the metastable helical states cannot be detected, as all of them, save the equilibrium state with $p=1$, are destabilized by boundary effects. In our opinion, it constitutes an exciting challenge to fabricate the composite magnet studied in this work and verify whether the metastable helical states actually exist, or, in other words, whether the dynamical topological protection is indeed effective. We hope the results presented in this paper will encourage work in this direction.

\begin{acknowledgments}
Grants No. PID2025-169885NB-I00, funded by MCIN/AEI/10.13039/501100011033, and E11\_26R/M4, funded by Diputaci\'on General de Arag\'on, supported this work.  
\end{acknowledgments}
	
\appendix

\section{Equations of stable static states \label{app:static}}

To study static states we find it convenient to parametrize the magnetization $\vm$ with two angles, $\theta$ and $\varphi$, introducing at each point $z$ a local right-handed orthonormal triad $\{\ve_1,\ve_2,\vm\}$, given by
\begin{gather}
\vm = \sin\theta\cos\varphi\,\hx + \sin\theta\sin\varphi\,\hy + \cos\theta\,\hz,
\\
\ve_1 = \cos\theta\cos\varphi\,\hx + \cos\theta\sin\varphi\,\hy - \sin\theta\,\hz,
\\
\ve_2 = -\sin\varphi\,\hx + \cos\varphi\,\hy.
\end{gather}
Recall that $\hz$ coincides with the chiral axis of the monoaxial chiral magnet. For other notation see Sec. \ref{sec:model}.

In terms of $\theta$ and $\varphi$ the static equations in the chiral magnet region, $-L_0<z<L_0$, have the form
\begin{gather}
\begin{gathered}
\theta^{\prime\prime} -\sin\theta\cos\theta\Big((\varphi^\prime-q_c)^2-h_c q_c^2\Big) 
- hq_c^2\sin\theta 
\\
 + vq_c \sin\theta\,\varphi^\prime - v\beta q_c\,\theta^{\,\prime} = 0,
\end{gathered}
\label{eq:static1}
\\
\varphi^{\prime\prime} + 2\cot\theta\,\theta^\prime(\varphi^\prime -q_c)  
- vq_c\frac{\theta^{\,\prime}}{\sin\theta} -v \beta q_c\,\varphi^\prime = 0,
\label{eq:static2}
\end{gather}
while in the ferromagnetic slabs, $L_0<|z|<L$, we have
\begin{gather}
\begin{gathered}
\theta^{\prime\prime} -\sin\theta\cos\theta\Big(\varphi^{\prime\,2} - q_c^2\kappa_m - q_c^2\kappa_u\cos^2\varphi\Big)  
\\
- \frac{h\mu q_c^2}{\rho}\sin\theta  + \frac{v\mu^2 q_c}{\rho}\sin\theta\,\varphi^\prime - \frac{v\beta\mu^2 q_c}{\rho}\,\theta^\prime = 0,
\end{gathered}
\label{eq:static3}
\\
\begin{gathered}
\varphi^{\prime\prime} + 2\cot\theta\,\theta^\prime\varphi^\prime - q_c^2\kappa_u\sin\varphi\cos\varphi
- \frac{v\mu^2 q_c}{\rho}\,\frac{\theta^\prime}{\sin\theta} 
\\
- \frac{v\beta\mu^2 q_c}{\rho}\,\varphi^\prime = 0.
\end{gathered}
\label{eq:static4}
\end{gather}

Static states are only relevant if they are stable. To analyze the linear stability we introduce a small perturbation of the static magnetization, $\vm_s$, in terms of the two component field $\xi=(\xi_1,\xi_2)^T$, so that
\be
\vm = \sqrt{1-\xi^2}\,\vm_s + \xi_1\,\ve_1+\xi_2\,\ve_2.
\ee
Now $\{\ve_1,\ve_2,\vm_s\}$ is an orthonormal triad.
Since $\vm_s$ is a solution to the static equation, to linear order in $\xi$ the LLG equation has the form
\be
\partial_t\xi = \omega_c (J-\alpha I)K\xi, \quad J = \begin{pmatrix} 0 & -I \\ I & 0\end{pmatrix}, \label{eq:LLG_linear}
\ee
where, for $i,j=1,2$,
$K_{ij} = T\delta_{ij} + A\,\epsilon_{ij} + S_{ij}$.

The static state of magnetization $\vm_s$ will be (linearly) stable if the spectrum of the operator $(J-\alpha I)K$ lies on the closed left complex half-plane (the set of complex numbers with non positive real part). 

At each interior point $z$ (that is, excluding the interface and boundary points) the operators act as
\begin{gather}
T =- \frac{a}{q_c^2}\,\frac{d^2}{dz^2} - \frac{a_m v\beta}{q_c}\,\frac{d}{dz} - w_s + \frac{a_m v}{q_c}\,\he_1\cdot\he_2^{\,\prime},
\\[8pt]
\begin{gathered}
A =\Big(-\frac{2a}{q_c^2}\,\he_1\cdot\he_2^\prime-\frac{2\chi_c}{q_c}\,\hz\cdot\vm_s+\frac{a_m v}{q_c}\Big)\,\frac{d}{dz} 
\\
-\frac{a}{q_c^2}\,(\he_1\cdot\he_2^\prime)^\prime-\frac{\chi_c}{q_c}\,(\hz\cdot\vm_s)^\prime-\frac{a_m v\beta}{q_c}\he_1\cdot\he_2^\prime,
\end{gathered}
\\[8pt]
\begin{gathered}
S_{ij} = \frac{a}{q_c^2}\,\he_i^{\,\prime}\cdot\he_j^{\,\prime}
-\frac{\chi_c}{q_c}\hz\cdot(\he_i\times\he_j^\prime+\he_j\times\he_i^\prime) 
\\
-a_z(\hz\cdot\he_i)(\hz\cdot\he_j)-\chi_u\kappa_u\frac{\rho}{\mu}(\hx\cdot\he_i)(\hx\cdot\he_j),
\end{gathered}
\end{gather}
where $a=a_s/a_m$ and $a_z = \kappa_c\chi_c - (\kappa_m\rho/\mu)\chi_u$, and
\be
\begin{gathered}
w _s =\frac{a}{q_c^2}\vm_s^{\prime\,2}+\frac{2\chi_c}{q_c}\vm_s\cdot(\hz\times\vm_s^\prime)
\\
-\chi_u\kappa_u\frac{\rho}{\mu}(\hx\cdot\vm_s)^2-a_z(\hz\cdot\vm_s)^2 - \vh\cdot\vm_s.
\end{gathered}
\ee
The linearized LLG equation (\ref{eq:LLG_linear}) has to be satisfied at each interior point of the materials. The vector fields $\vm$ and $\vc$ are continuous if and only if $\xi_\alpha$ and
\be
d_\alpha = a_s\xi_\alpha^\prime + \sum_{\beta=1}^2\big(a_s\ve_\alpha\cdot\ve^{\,\prime}_\beta + q_c\chi_c(\hz\cdot\vm_s)\epsilon_{\alpha\beta}\big)\xi_\beta
\ee
are continuous for $\alpha=1,2$. In the above expression $\epsilon_{\alpha\beta}$ is the totally antisymmetric tensor in two dimensions. Hence, $\xi_\alpha$ and $d_\alpha$ have to be continuous across the internal intefaces ($z=\pm L_0$), and $d_\alpha$ has to vanish at the external boundaries ($z=\pm L$).

\section{Numerical methods}

We have developed two algorithms to obtain the numerical approximations to the solutions of the time-dependent LLG equation and the static equations. Both are based on finite difference discretizations. The algorithms have been implemented in two computational codes which are called the \textit{dynamic code} and the \textit{static code}. Below we briefly describe the two algorithms.

\textit{Dynamical algorithm}. It provides a numerical solution to the time-dependent LLG equation as an evolution problem, starting from a given initial condition. A regular grid is introduced in the interval $[-L,L]$such that the interface points $z=\pm L_0$ belong to the grid. This requires $L$ and $L_{0}$ to be commensurate, a restriction that causes no practical loss of generality. The interior grid points are those with $z\neq \pm L_0,\pm L$, while the boundary and interface points are located at $z=\pm L$ and $z=\pm L_0$, respectively. The initial condition provides the magnetization at the interior grid points. The magnetization at the interface points is determined by the continuity of $\vc(z)$ at $z=\pm L_0$. The lateral derivatives of $\vm$ are computed using second-order forward and backward finite-difference schemes on three-point stencils. The magnetization at the boundary points is computed similarly by setting $\vc(\pm L) = 0$. The effective field (including the torque delivered by the current) is computed only at the interior grid points using second-order central finite-difference schemes. Given the effective field, the magnetization at the interior grid points is advanced one step in time using the explicit RK4 method. Subsequently, the magnetization is renormalized to enforce a unit vector field. Once the magnetization at the interior grid points is advanced, the values at the boundary and interface points are properly updated. Since explicit methods for PDEs are inherently unstable unless the spatial and time steps satisfy a strict stability condition, we ensure numerical stability by choosing a sufficiently small time step.

\textit{Static algorithm}. It provides a numerical solution to the static LLG equations given by Eqs. (\ref{eq:static1})-(\ref{eq:static4}). The two second-order equations for $\theta$ and $\varphi$ are equivalent to a system of four first-order equations, obtained by defining the new variables $w_\theta=\theta^\prime$ and $w_\varphi=\varphi^\prime$. This system is discretized using a first-order forward-difference scheme, on a regular grid identical to the grid introduced for the dynamical algorithm, and is enforced only at the interior grid points.
On the boundary points, the boundary conditions $d_\alpha=0$ are enforced. At the interface points, $\theta$ and $\varphi$ are continuous (provided $\sin\theta\neq 0$) but $w_\theta$ and $w_\varphi$ may be discontinuous. Hence, at the interface points both $w_\theta$ and $w_\varphi$ have two values, which represent the one-sided limits of the corresponding functions, and the discretized equations enforce the continuity of $d_\alpha$. The discretized system is solved using the Newton--Raphson method with overrelaxation. 

\bibliographystyle{unsrt}
\bibliography{references}

 \end{document}